\documentclass[12pt]{article}
\usepackage{times}
\usepackage{geometry}
\usepackage{floatflt}
\usepackage[utf8]{inputenc}
\usepackage{enumitem,amssymb}
\usepackage{ragged2e}
\usepackage{multicol}
\newlist{thematic}{itemize}{8}
\setlist[thematic]{label=$\square$}
\usepackage{pifont}
\usepackage{graphicx}
\usepackage{tabularx}
\usepackage{sidecap}
\usepackage[utf8]{inputenc}
\usepackage[vietnamese,english]{babel}
\usepackage[dvipsnames,svgnames,x11names]{xcolor}
\usepackage{jheppub}   
\usepackage{sectsty}

\definecolor{DarkGreen}{rgb}{0.0, 0.3, 0.0}
\definecolor{purple}{rgb}{0.5, 0.0, 0.5}
\definecolor{red}{rgb}{1, 0.0, 0.0}
\definecolor{green}{rgb}{0, 1.0, 0.0}

\def\3he{$^3{\rm He}$}

\def\lsim{\mathrel{\lower2.5pt\vbox{\lineskip=0pt\baselineskip=0pt
           \hbox{$<$}\hbox{$\sim$}}}}

\def\gsim{\mathrel{\lower2.5pt\vbox{\lineskip=0pt\baselineskip=0pt
           \hbox{$>$}\hbox{$\sim$}}}}

\begin{document}
\raggedright
\huge
Astro2020 Science White Paper \linebreak

Science from an Ultra-Deep, High-Resolution Millimeter-Wave Survey\linebreak
\normalsize

\noindent \textbf{Thematic Areas:} \linebreak
Primary Area: Cosmology and Fundamental Physics \linebreak
Secondary Area: Galaxy Evolution \linebreak

  
\textbf{Principal Author:}

Name: Neelima Sehgal
 \linebreak						
Institution:  Stony Brook University \& Flatiron Institute
 \linebreak
Email:  neelima.sehgal@stonybrook.edu
 \linebreak
 Phone:  631-632-8229
 \linebreak
 
\textbf{Co-authors:} 
\foreignlanguage{vietnamese}{Hồ Nam Nguyễn}$^1$,
Joel Meyers$^2$,
Moritz Munchmeyer$^1$,
Tony Mroczkowski$^3$, 
Luca Di Mascolo$^4$, 
Eric Baxter$^5$, 
Francis-Yan Cyr-Racine$^{6,7}$, 
Mathew Madhavacheril$^8$,
Benjamin Beringue$^9$,
Gil Holder$^{10}$,
Daisuke Nagai$^{11}$,
Simon Dicker$^5$,
Cora Dvorkin$^6$,
Simone Ferraro$^{12}$,
George M. Fuller$^{13}$,
Vera Gluscevic$^{14}$,
Dongwon Han$^{15}$, 
Bhuvnesh Jain$^5$,
Bradley Johnson$^{16}$, 
Pamela Klaassen$^{17}$, 
Daan Meerburg$^9$,
Pavel Motloch$^{18}$, 
David N Spergel$^{19,8}$, 
Alexander van Engelen$^{18}$ 
\linebreak

\textbf{Endorsers:} 
Peter Adshead$^{10}$, 
Robert Armstrong$^{20}$, 
Carlo Baccigalupi$^{21}$,
Darcy Barron$^7$, 
Kaustuv Basu$^{22}$, 
Bradford Benson$^{23,24}$, 
Florian Beutler$^{25}$, 
J. Richard Bond$^{18}$, 
Julian Borrill$^{12}$, 
Erminia Calabrese$^{26}$, 
Omar Darwish$^9$, 
S. Lucas Denny$^{27}$, 
Kelly A. Douglass$^{28}$, 
Tom Essinger-Hileman$^{29}$, 
Simon Foreman$^{18}$, 
David Frayer$^{30}$, 
Martina Gerbino$^{31}$, 
Satya Gontcho A Gontcho$^{32}$, 
Evan B. Grohs$^{33}$,
Nikhel Gupta$^{34}$, 
J. Colin Hill$^{35,19}$,
Christopher M. Hirata$^{36}$,
Selim Hotinli$^{37}$,
Matthew C. Johnson$^{38,1}$,
Marc Kamionkowski$^{39}$, 
Ely D.~Kovetz$^{40}$, 
Erwin T.\ Lau$^{41}$, 
Michele Liguori$^{46}$,
Toshiya Namikawa$^9$, 
Laura Newburgh$^{11}$,
Bruce Partridge$^{42}$, 
Francesco Piacentni$^{43}$, 
Benjamin Rose$^{44}$, 
Graziano Rossi$^{45}$, 
Benjamin Saliwanchik$^{11}$, 
Emmanuel Schaan$^{12,33}$,
Huanyuan Shan$^{47}$, 
Sara Simon$^{48}$,
An\v{z}e Slosar$^{49}$, 
Eric R. Switzer$^{50}$, 
Hy Trac$^{51}$,
Weishuang Xu$^6$, 
Matias Zaldarriaga$^{35}$, 
Michael Zemcov$^{52}$ 
\justify

\vspace{3mm}
\textbf{Abstract:}
    Opening up a new window of millimeter-wave observations that span frequency bands in the range of 30 to 500 GHz, survey half the sky, and are both an order of magnitude deeper (about $0.5~\mu$K-arcmin) and of higher-resolution (about 10 arcseconds) than currently funded surveys would yield an enormous gain in understanding of both fundamental physics and astrophysics.  In particular, such a survey would allow for major advances in measuring the distribution of dark matter and gas on small-scales, and yield needed insight on 1.)~dark matter particle properties, 2.)~the evolution of gas and galaxies, 3.)~new light particle species, 4.)~the epoch of inflation, and 5.)~the census of bodies orbiting in the outer Solar System.

\thispagestyle{empty}
\pagebreak
\setcounter{page}{1}

\section{Introduction}

\vspace{-2mm}
Opening up a new window of millimeter-wave survey observations that are both deeper and of higher-resolution than previous surveys would yield an enormous gain in understanding of both fundamental physics and astrophysics.   
The major advances enabled by a deep (about $0.5~\mu$K-arcmin), high-resolution (about 10 arcsecond), millimeter-wave survey are: i)~the use of gravitational lensing of the primordial microwave background to map out the distribution of matter on small scales ($k\sim10~h$Mpc$^{-1}$), and ii)~the measurement of the thermal and kinetic Sunyaev-Zel'dovich effects (tSZ and kSZ) on small scales to map the gas density and gas pressure {\it{profiles}} of galaxy clusters and groups.  In addition, such a survey, covering half the sky, would allow us to cross critical thresholds in fundamental physics: i.)~ruling out or detecting any new, light, thermal particles, which could potentially be the dark matter, and ii)~testing a wide class of multi-field models that could explain an epoch of inflation in the early Universe.  Lastly, such a survey enables the detection of an Earth-size planet thousands of AU from the Sun, and more broadly provides a census of asteroids and planetary bodies in the outer Solar System.  This survey would also enable the detection of exo-Oort clouds around other solar systems, shedding light on planet formation.

\vspace{-2mm}
\section{Dark Matter Distribution and Properties}

\vspace{-1mm}
{\bf{What is the distribution of matter on small scales?}}

\vspace{0.2cm}
\noindent{\bf{What are the particle properties of dark matter?}}\\

\vspace{-2mm} 
There is compelling evidence for non-baryonic dark matter~\cite{Zwicky1937,Rubin1970,Ostriker1974,Fabricant1980,Bahcall1995,Clowe2006,Planck2018}, however, its nature remains elusive.  Direct and indirect dark-matter searches may set tight limits in the next decade, but still turn up empty-handed in terms of detecting a new particle.  One may then ask what is the next best avenue to probe dark-matter properties.  The only place we have directly detected the impact of dark matter is through its gravitational interactions.  Thus it makes sense to explore this direction for further insight into dark-matter particle properties.  In particular, well-motivated models of dark matter that differ from the standard cold, collisionless model (CDM), predict different distributions of matter on small scales~\cite{Colin2000,Bode2001,Boehm:2001hm,Boehm:2004th,Viel2005,Turner1983,Press1990,Sin1994,Hu2000,Goodman2000,Peebles_2000,Amendola2006,Schive2014,Marsh2016,Carlson1992,Spergel2000,Vogelsberger2012,Dvorkin:2013cea,Fry2015,Elbert2015,Kaplinghat2016,Kamada2017,Gluscevic2018,Boddy2018a,Li2018,Boddy2018b,Tulin2018,Khoury2016,Vogelsberger2016,Cyr-Racine:2013ab,Cyr-Racine:2015ihg,Schewtschenko:2014fca,Krall:2017xcw,Xu:2018efh}. However, observations of the small-scale matter distribution to date are challenging because they often infer the matter distribution through baryonic tracers of the matter~\cite{Koposov2015,Drlica-Wagner2015,Menci2017,Mesinger2005,deSouza2013,Moore1999,Johnston2016,Carlberg2009,Erkal2015,Bovy2014,Cen1994,Hernquist1996,Croft1999,Hui1999,Viel2013,Baur2016,Irsic2017}, which may not trace the dark matter reliably~\cite{Sawala2016,Oman2015,Hui2017}.  While very promising, alternative techniques using strong gravitational lensing to find low-mass dark-matter halos \cite{Dalal2002,Koopmans_2005,Vegetti:2009aa,Vegetti_2010_1,Vegetti_2010_2,Vegetti2012,Hezaveh:2012ai,Vegetti2014,Hezaveh2016a,Ritondale:2018cvp} face the challenge of disentangling the complex structure of the background source from the substructure signal; using strong lensing to measure the matter power spectrum faces a similar challenge~\cite{Hezaveh2016b,Daylan:2017kfh,Bayer:2018vhy,DiazRivero2018,Chatterjee2018,Cyr-Racine:2018htu,Brennan2018,Rivero:2018bcd}. In contrast, {\it{measuring the small-scale matter power spectrum from weak gravitational lensing using the cosmic microwave background (CMB) as a backlight}} suffers from no such degeneracies, since we have detailed knowledge of the intrinsic properties of the source~\cite{Nguyen2019}. This may provide a powerful complementary probe of dark-matter physics~\cite{Nguyen2019}.    

\begin{SCfigure}[1.4][t]
\centering
\includegraphics[width=0.48\textwidth,height=6.6cm]{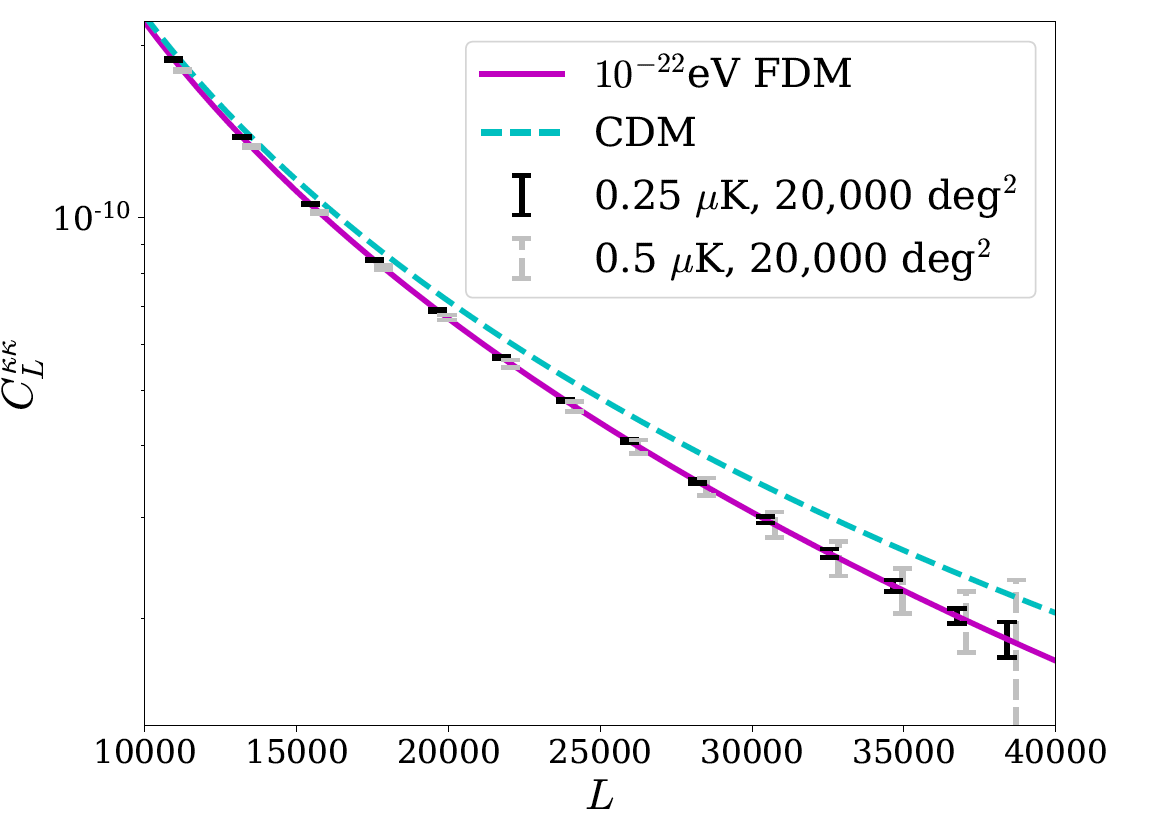}
\caption{CMB lensing power spectrum for an $m \sim 10^{-22}$ eV FDM model and a CDM model.  Here the black and silver error bars correspond to $0.25 \mu$K-arcmin and $0.5 \mu$K-arcmin CMB noise in temperature, respectively. For both sets of error bars, 10'' resolution observations covering a $50\%$ sky fraction is assumed.  Given the black and silver error bars, one can distinguish between FDM and CDM with a signal-to-noise ratio of about 24 and 11, respectively, in the absence of baryons. No kSZ has been included.}
\label{fig:dm}  
\vspace{-4mm}
\end{SCfigure}

As shown in Figure~\ref{fig:dm}, by measuring the small-scale matter power spectrum, one can distinguish between CDM and a dark matter model that alters the matter distribution in a way that differs from the CDM prediction~\cite{Nguyen2019}.  Such a measurement has the potential to determine both dark-matter particle properties and the effect of baryons on the small-scale matter distribution; both of which can affect the shape of the small-scale matter power spectrum, albeit in potentially different ways~\cite{Nguyen2019,vanDaalen2011,Brooks2013,Brooks2014,Natarajan2014,Schneider2018}.  Recently, it has been shown, by comparing the output of multiple hydrodynamic simulations with different baryonic prescriptions, that there may be a finite set of ways that baryons can alter the matter power spectrum; moreover, these changes to the matter power spectrum may depend on just a few free parameters~\cite{Schneider2018} that can potentially be distinguished from dark matter model parameters.  We note that such a measurement of the small-scale matter power spectrum, in any case, will constrain both models of dark matter and baryonic physics. 

\vspace{-4mm}
\begin{center}
\begin{table}[b]
\centering
\begin{tabularx}{0.77\linewidth}{|c|c|c|c|c|}
	\hline
	\hline
 	Sky fraction & Noise at 150 GHz & \multicolumn{3}{c|}{Dark matter signal-to-noise ratio}   \\
    (f\textsubscript{sky})& ($\mu$K-arcmin) & no kSZ & reion kSZ & reion+late kSZ \\
    \hline    
    0.5 & 0.5 & 11 & 9 & 5\\ 
    0.5 & 0.25 & 24 & 19 & 7\\
    \hline
	0.25 & 0.5 & 8 & 7 & 4\\
	0.25 & 0.25 & 17 & 13 & 5\\
	\hline
	0.1 & 0.5 & 5 & 4 & 2\\
	0.1 & 0.25 & 11 & 9 & 3\\
	\hline 
\end{tabularx}
\caption{Significance with which an $m\sim10^{-22}$~eV FDM (or 1 keV WDM) model can be distinguished from a CDM model, based on observations of CMB lensing.  Here we assume 10'' resolution and vary observed sky fraction, noise levels in temperature, and residual kSZ.}
\label{tab:dm}
\end{table}
\vspace{-5mm}
\end{center}

\vspace{-2mm}
In Table~\ref{tab:dm}, we show forecasts of how well one can potentially distinguish between fuzzy or warm dark matter models (FDM or WDM) and CDM, in the absence of baryonic effects.  We show this with differing levels of residual kSZ, which is a frequency-independent foreground that arises when CMB photons are Doppler boosted after scattering off ionized gas with a bulk motion.  This kSZ signal arises from scattering off the ionized gas bound in dark-matter halos (late-time kSZ), and the ionized gas present during the epoch of reionization when the first stars were forming (reionization kSZ).   From Table~\ref{tab:dm}, we see that the reionization kSZ has a modest impact on the signal-to-noise ratio, while we potentially can remove much of the late-time kSZ with techniques facilitated by an overlapping galaxy survey~\cite{Schaan2016,Smith2018}.  Frequency-dependent foregrounds can likely be removed with a combination of multi-frequency channels and high-resolution observations, the latter of which allow extragalactic foregrounds to be resolved and subtracted from survey maps.   

\section{Baryonic Physics and Galaxy Evolution}
    
\vspace{-1mm}
{\bf{What is the evolution of gas in and around dark matter halos?}}\\

\begin{figure}
\centering
\vspace{-5mm}
\includegraphics[width=0.86\textwidth]{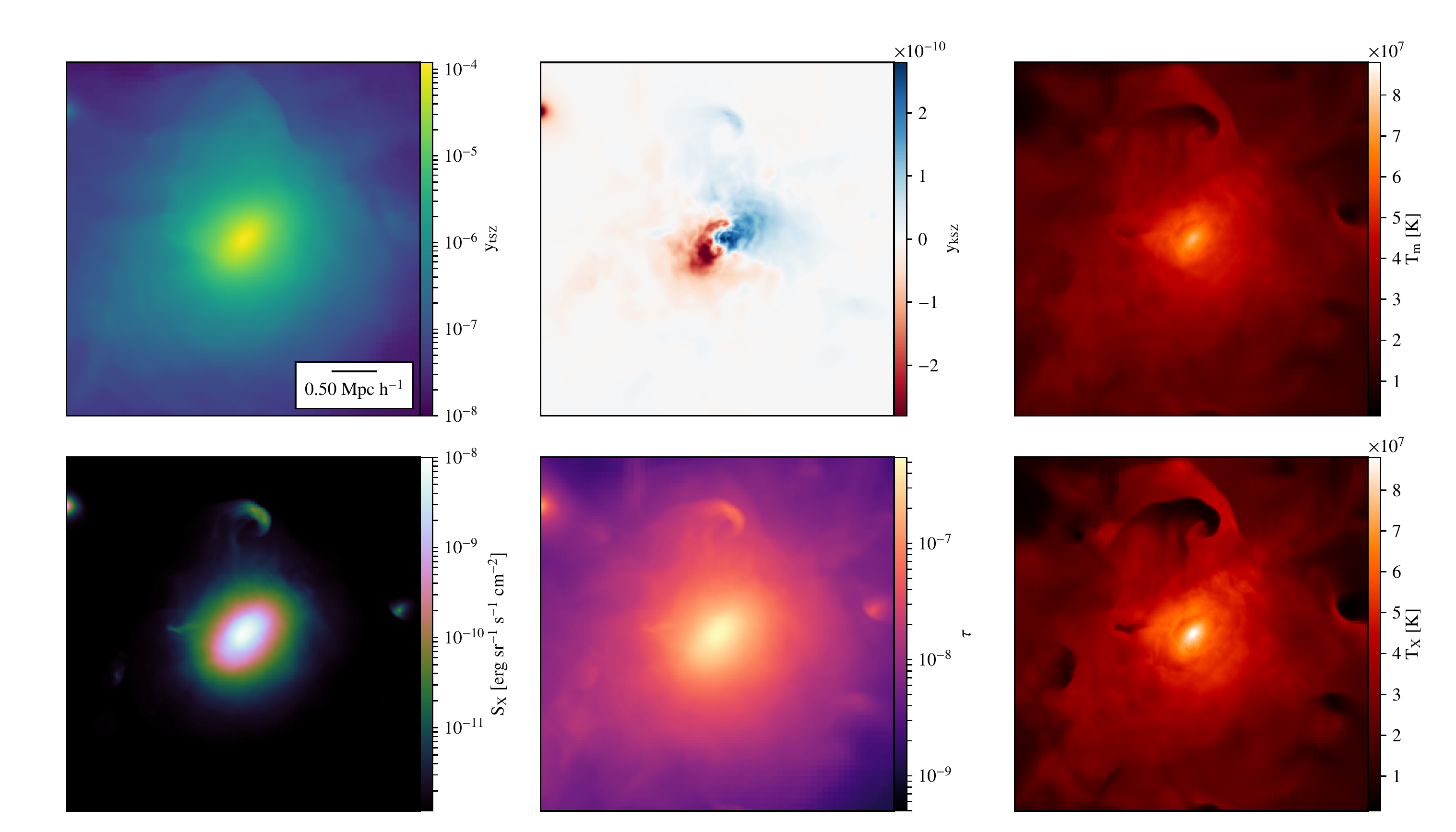}
\vspace{-4mm}
\caption{The SZ effects of a merging galaxy cluster extracted from the {\it Omega500} cosmological hydrodynamical simulation \cite{Nelson2014a}, with some X-ray quantities for comparison. 
Shown quantities are the tSZ signal (top-left), kSZ signal (top-middle), projected mass-weighted temperature (top-right), X-ray surface brightness (bottom-left), electron opacity (bottom-middle), and projected X-ray temperature (bottom-right). 
{\bf{Combining low-noise, high-resolution tSZ and kSZ measurements allows one to separate the pressure, density, velocity, and temperature of the gas}}~\cite{Knox2004,Sehgal2005,Battaglia2017}. }
\label{fig:SZ}  
\vspace{-3mm}
\end{figure}

\vspace{-3mm}
A high-resolution, low-noise millimeter-wave survey would open a new window on tSZ and kSZ measurements \cite[][for a recent review]{Mroczkowski2019}. tSZ measurements probe the {\it{thermal pressure}} of ionized gas in and around galaxy clusters and groups~\cite{Sunyaev1970}. kSZ measurements probe the {\it{gas momentum density}}~\cite{Sunyaev1972}.  The combination of kSZ and tSZ measurements at three or more frequencies, with low-noise and high-resolution, would allow one to measure the gas density, bulk velocity, and temperature separately \cite{Knox2004,Sehgal2005,Battaglia2017}. The advantage of SZ over X-ray measurements is that the SZ signal is proportional to the gas density (not density squared) and the flux of the signal is redshift independent; thus SZ measurements are a powerful probe of the gas in galaxy cluster outskirts, in low-mass halos, and in high-redshift halos.  

Over the past decade, the tSZ effect has also been used to {\it{find}} hundreds of previously-unknown clusters of galaxies out to $z \sim 2$~\cite{Marriage2011,Vanderlinde2010,Planck2014}.  Higher resolution and better sensitivity will push this to lower-mass and higher-redshift systems, allowing direct imaging of many systems where X-ray observations would require exceptionally long integration times. Stacking the tSZ detections will allow one to probe the gas physics in and around halos out to $z\sim 2$ and with masses below $10^{12}$ M$_\odot$. 
The gas in halos reflects the impact of feedback and merging processes and serves as a reservoir of star formation.  For example, as shown in Figure~\ref{fig:SZ}, tSZ and kSZ measurements probe gas accretion, non-thermal and non-equilibrium processes, AGN feedback, and splashback and shock radii~\cite{Nagai2011,Nelson2014b,Lau2015,Avestruz2015}.  Therefore, the science gain of such measurements is a deeper understanding of galaxy cluster astrophysics, the physics of the intergalactic and circumgalactic medium, and galaxy evolution.

\begin{center}
\begin{table}[t]
\centering
\begin{tabularx}{0.838\linewidth}{|c|c|c|c|}
	\hline
	\hline
 	Sky fraction & Noise at 150 GHz & \multicolumn{2}{c|}{Constraint on new particle species ($\sigma({N_{\rm{eff}}})$)}   \\
    (f\textsubscript{sky})& ($\mu$K-arcmin) & with kSZ foregrounds & no foregrounds \\
    \hline    
    0.5 & 0.5 & 0.019 & 0.014\\
    0.5 & 0.25 &  0.018 & 0.013\\
    \hline
	0.25 & 0.5 & 0.027 & 0.019\\
	0.25 & 0.25 & 0.025 & 0.018\\
	\hline 
\end{tabularx}
\caption{Constraint on the density of light relics, $N_{\rm{eff}}$.  We assume 10'' resolution and vary the noise and sky area. We show constraints with and without including the kSZ foregrounds for temperature maps. {\it{Planck}} and expected Simons Observatory data are included.  {\bf{Note that the critical threshold of $\sigma({N_{\rm{eff}}}) = 0.027$
is the minimum change to $N_{\rm{eff}}$ created by {\it{any}} new light particle species in thermal equilibrium with standard model particles in the early Universe}}~\cite{CMBS4SB}.}
\label{tab:neff}
\end{table}
\end{center}

\vspace{-20mm}
\section{Thermal History of The Early Universe}

\vspace{-1mm}
{\bf{Do new light particle species exist that were in thermal equilibrium with standard model particles during the early Universe?}}\\

\vspace{-2mm}
Any light particles that were relativistic when the CMB formed (i.e., masses less than about 0.1 eV) and that were in thermal equilibrium with standard model particles at any time in the early Universe leave an imprint on the CMB power spectrum at small angular scales.  
The critical threshold for the density of light relics, $N_{\rm{eff}}$, is $\sigma({N_{\rm{eff}}}) = 0.027$,
which represents the minimum change to $N_{\rm{eff}}$ created by {\it{any}} new light species in thermal equilibrium with standard model particles in the early Universe (assuming no significant dilution by new states beyond the standard model particle content)~\cite{CMBS4SB}.  This includes light species that were in thermal equilibrium with the standard model right after the Big Bang, probing back to about $10^{-30}$ seconds after the creation of the Universe.  These particles may well be too weakly interacting to be seen with significance in Earth-based experiments; thus probing them with cosmology may be the only avenue.  Many dark matter models, including many axion models, predict new light thermal relics~\cite{Baumann:2016wac,Green:2017ybv}.  Thus ruling out, or detecting, the existence of such a particle would yield insight on dark matter particle properties.  

With an ultra-deep, high-resolution millimeter-wave survey, one could cross the critical threshold of $\sigma({N_{\rm{eff}}}) = 0.027$.  The ability to cross this threshold comes from i)~surveying a wide sky area, ii)~having low noise, and iii)~probing small angular scales ($\ell_{\rm{max}} \gtrsim 5000$).
In Table~\ref{tab:neff}, we show the constraints on $N_{\rm{eff}}$ for different noise levels and sky areas, for a $\Lambda {\rm{CDM}}+N_{\rm{eff}}+\sum m_{\nu}$ model and assuming BBN consistency.  We also show constraints with and without including the kSZ foregrounds in temperature maps; this foreground can potentially be reduced with an overlapping galaxy survey. 
We assume 10 arcsecond resolution and use lensed CMB spectra.   We note that delensing the CMB spectra can modestly improve the $N_{\rm{eff}}$ constraints~\cite{Baumann:2015rya,Green:2016cjr}, and including the effect of Rayleigh scattering can potentially have a significant impact depending on the noise levels of the higher-frequency channels (e.g.~15-30\% improvement with 5/10 $\mu$K-arcmin noise at 400/500 GHz)~\cite{Lewis2013,Alipour:2014dza}.  
From Table~\ref{tab:neff}, in a scenario where foregrounds are effectively removed (via multi-frequency cleaning, resolving and subtracting extragalactic sources, and removing kSZ from the maps), {\it{such a survey can potentially rule out or find evidence for new light thermal particles with $95\%$ confidence level.}}

\vspace{1mm}
\section{Signatures from the Newborn Universe}

\vspace{-3mm}
{\bf{Do primordial gravitational waves from an epoch of inflation exist?}}

\vspace{0.1cm}
{\noindent\bf{If inflation happened, did it arise from multiple or a single new field?}}\\

\vspace{-4mm}
In order to detect primordial gravitational waves from inflation via their B-mode fluctuations, using a ground-based experiment, it is necessary to subtract the B-mode fluctuations due to gravitational lensing of E-modes.  To reach a target for the tensor-to-scalar ratio, $r$, of $\sigma(r)<5 \times 10^{-4}$, requires removal of at least $90\%$ of the lensing B-mode power, i.e. $A_{\rm{lens}}\leq 0.1$~\cite{CMBS4SB}.  This $\sigma(r)$ target, in the absence of a detection, would rule out or disfavor all inflation models that naturally explain the observed value of the scalar spectral index, and which have a characteristic scale in field space larger than the Planck scale~\cite{CMBS4SB}.  The third column of Table~\ref{tab:inflation} shows that such an $A_{\rm{lens}}$ target is well within reach.  In addition, as shown in the fourth column, maps of the kSZ signal cross correlated with an overlapping galaxy survey such as LSST, can constrain primordial non-Gaussianity, crossing the critical threshold of $\sigma(f_{\rm{NL}}) < 1$ that would rule out a wide class of multi-field inflation models~\cite{Alvarez:2014vva,Smith2018,Munchmeyer2018,Deutsch2018}.
\begin{center}
\begin{table}[t]
\centering
\begin{tabularx}{0.938\linewidth}{|c|c|c|c|}
	\hline
	\hline
 	Sky fraction & Noise & \multicolumn{2}{c|}{De-lensing ability and non-Gaussianity constraints}   \\
    (f\textsubscript{sky})& ($\mu$K-arcmin) & Resid.\ lensing B-mode ($A_{\rm{lens}}$) & Non-Gaussianity ($\sigma(f_{\rm{NL}})$) \\
    \hline    
    0.5 & 0.5 & 0.1 & 0.26 \\ 
    0.5 & 0.25 & 0.05 & 0.26 \\
    \hline
	0.25 & 0.5 & 0.1 & 0.40 \\
	0.25 & 0.25 & 0.05 & 0.40 \\
	\hline 
\end{tabularx}
\vspace{-2mm}
\caption{{\it{Third column:}} Expected residual lensing B-mode power after delensing ($A_{\rm{lens}}$).  {\bf{Removing the lensing B-mode signal is critical to detecting primordial gravitational waves from inflation; for example, $A_{\rm{lens}}\leq 0.1$ is necessary to achieve $\sigma(r)=5\times 10^{-4}$.}} 
Note we adopt a conservative $l_{\rm min}=1000$ cut for B-modes; $l_{\rm min}=100$ for B-modes decreases $A_{\rm{lens}}$ by $30\%$.  
{\it{Fourth column:}} Constraints on primordial non-Gaussianity from a kSZ survey overlapping with LSST~\cite{Munchmeyer2018}.  These constraints are galaxy-shot-noise limited, yielding minimal gain from lowering the noise below $0.5 \mu$K-arcmin.   {\bf{Achieving $\sigma(f_{\rm{NL}}) < 1$ can distinguish between single and multi-field inflation models.}} We assume 10'' resolution.}
\label{tab:inflation}
\vspace{-2mm}
\end{table}
\end{center}

\vspace{-15mm}
\section{Planetary and Exo-Solar System Studies}

\vspace{-3mm}
{\bf{What is the census of bodies in the outer Solar System, and around other stars?}}\\

\vspace{-4mm}
Millimeter-wave surveys can open a new discovery space of bodies in the outer Solar System, and planetary systems around other stars.  High-resolution, low-noise millimeter-wave surveys could detect undiscovered Solar System planets via their thermal flux and parallactic motion~\cite{Cowan2016}.  In the outer Solar System, the main source of heating for planets is internal.  Optical observations, in contrast, are sensitive to the bodies' reflected light from the Sun.  Since the flux from reflected light falls faster with distance than directly sourced emission, millimeter-wave surveys have an advantage over optical surveys in finding objects in the far Solar System~\cite{Baxter2018a}.  A deep, high-resolution millimeter-wave survey could detect, for example, Earth-sized planets at thousands of AU from the Sun.  Such a survey, combined with optical measurements, would allow large population studies of the sizes and albedos of these objects~\cite{Gerdes:2017}. 
Deep millimeter-wave surveys can also enable the detection of {\it{exo-Oort clouds around other stars}}~\cite{Baxter2018b}, opening a new window into the study of planetary systems.

\clearpage
\section*{Affiliations}

\begin{multicols}{2}
\noindent $^1$ Perimeter Institute \\
$^2$ Southern Methodist University \\
$^3$ European Southern Observatory \\
$^4$ Max Planck Institute for Astrophysics \\
$^5$ University of Pennsylvania \\
$^6$ Harvard University \\
$^7$ University of New Mexico \\
$^8$ Princeton University \\
$^9$ University of Cambridge \\
$^{10}$ University of Illinois at Urbana-Champaign \\
$^{11}$ Yale University \\
$^{12}$ Lawrence Berkeley National Laboratory \\
$^{13}$ University of California, San Diego \\
$^{14}$ University of Florida \\
$^{15}$ Stony Brook University \\
$^{16}$ Columbia University \\
$^{17}$ UK Astronomy Technology Centre \\
$^{18}$ CITA, University of Toronto \\
$^{19}$ Flatiron Institute \\
$^{20}$ Lawrence Livermore National Laboratory \\
$^{21}$ SISSA, Trieste \\
$^{22}$ University of Bonn \\
$^{23}$ Fermi National Accelerator Laboratory \\
$^{24}$ Kavli Institute for Cosmological Physics \\
$^{25}$ University of Portsmouth \\
$^{26}$ Cardiff University \\
$^{27}$ Florida State University \\
$^{28}$ University of Rochester \\
$^{29}$ Goddard Space Flight Center \\
$^{30}$ Green Bank Observatory \\
$^{31}$ Argonne National Laboratory \\
$^{32}$ University of Rochester \\
$^{33}$ University of California, Berkeley \\
$^{34}$ The University of Melbourne \\
$^{35}$ Institute for Advanced Study \\
$^{36}$ The Ohio State University \\
$^{37}$ Imperial College, London \\
$^{38}$ York University \\
$^{39}$ Johns Hopkins University \\
$^{40}$ Ben-Gurion University \\
$^{41}$ University of Miami \\
$^{42}$ Haverford College \\
$^{43}$ Sapienza University of Rome \\
$^{44}$ Space Telescope Science Institute \\
$^{45}$ Sejong University \\
$^{46}$ University of Padova \\
$^{47}$ Shanghai Astronomical Observatory \\
$^{48}$ University of Michigan \\
$^{49}$ Brookhaven National Laboratory \\
$^{50}$ Goddard Space Flight Center \\
$^{51}$ Carnegie Mellon University \\
$^{52}$ Rochester Institute of Technology \\
\end{multicols}

\newpage

\bibliographystyle{unsrturltrunc6}
\bibliography{cmb-in-hd.bbl}

\end{document}